\documentclass[preprint,journal]{vgtc}       % preprint (journal style)

\ifpdf%                                % if we use pdflatex
  \pdfoutput=1\relax                   % create PDFs from pdfLaTeX
  \usepackage{graphicx}                % allow us to embed graphics files
  \DeclareGraphicsExtensions{.pdf,.png,.jpg,.jpeg} % for pdflatex we expect .pdf, .png, or .jpg files
\else%                                 % else we use pure latex
  \usepackage{graphicx}                % allow us to embed graphics files
  \DeclareGraphicsExtensions{.eps}     % for pure latex we expect eps files
\fi%

\graphicspath{{figures/}{pictures/}{images/}{./}} % where to search for the images

\usepackage{microtype}                 % use micro-typography (slightly more compact, better to read)
\PassOptionsToPackage{warn}{textcomp}  % to address font issues with \textrightarrow
\usepackage{textcomp}                  % use better special symbols
\usepackage{mathptmx}                  % use matching math font
\usepackage{times}                     % we use Times as the main font
\usepackage{cite}                      % needed to automatically sort the references
\usepackage{tabu}                      % only used for the table example
\usepackage{booktabs}
\usepackage{soul}
\usepackage{multirow}
\usepackage[dvipsnames]{xcolor}
\usepackage{colortbl}
\usepackage{color}
\usepackage{hyperref}

\definecolor{CindyX}{RGB}{232, 125, 114}

\definecolor{red}{HTML}{e85641}
\definecolor{blue}{HTML}{2656ad}

\newenvironment{tight_itemize}{\begin{itemize} \itemsep
-2.1pt}{\end{itemize}}

\newcommand{\pheading}[1]{\vspace{4px}\noindent\textbf{#1}}

\onlineid{1056}

\vgtccategory{Research}
\vgtcpapertype{please specify}

\title{Comparison Conundrum and the Chamber of Visualizations:\\An Exploration of How Language Influences Visual Design}

\author{Aimen Gaba, Vidya Setlur (\textit{Member}), Arjun Srinivasan, Jane Hoffswell, and Cindy Xiong}
\authorfooter{
\item Aimen Gaba and Cindy Xiong are with UMass Amherst. E-mail: \{agaba, yaxiong\}@umass.edu.
\item Vidya Setlur and Arjun Srinivasan are with Tableau Research. E-mail: \{vsetlur, arjunsrinivasan\}@tableau.com.
\item Jane Hoffswell is with Adobe Research. E-mail: jhoffs@adobe.com.
}

\shortauthortitle{Author 1 \MakeLowercase{\textit{et al.}}: Visualizations for Comparison Utterances}
\abstract{
The language for expressing comparisons is often complex and nuanced, making supporting natural language-based visual comparison a non-trivial task. To better understand how people reason about comparisons in natural language, we explore a design space of utterances for comparing data entities. We identified different parameters of comparison utterances that indicate \textbf{what} is being compared (i.e., data variables and attributes) as well as \textbf{how} these parameters are specified (i.e., explicitly or implicitly). We conducted a user study with sixteen data visualization experts and non-experts to investigate how they designed visualizations for comparisons in our design space. Based on the rich set of visualization techniques observed, we extracted key design features from the visualizations and synthesized them into a subset of sixteen representative visualization designs. We then conducted a follow-up study to validate user preferences for the sixteen representative visualizations corresponding to utterances in our design space. Findings from these studies suggest guidelines and future directions for designing natural language interfaces and recommendation tools to better support natural language comparisons in visual analytics.
}

\keywords{Comparative constructions, cardinality, explicit and implicit comparisons, natural language, intent, visual analysis.}

\CCScatlist{ % not used in journal version
 \CCScat{K.6.1}{Management of Computing and Information Systems}%
{Project and People Management}{Life Cycle};
 \CCScat{K.7.m}{The Computing Profession}{Miscellaneous}{Ethics}
}

\usepackage{xargs}      % Used for new commands with optional arguments
\usepackage{soul}       % Used for custom comments
\usepackage{color}      % Used for custom colors in comments
\usepackage{xspace}     % Used for abbreviation spacing
\usepackage{xpunctuate} % Used for abbreviation spacing

\newcommand{\myquote}[1]{\emph{``#1''}}                         % Italic text in quotation marks
\definecolor{oneToOneColor}{HTML}{08A4BD}
\definecolor{oneToNColor}{HTML}{D5A021}
\definecolor{nToMColor}{HTML}{498756}
\definecolor{nColor}{HTML}{9B5094}

\definecolor{cardGreen}{RGB}{165,201,148}
\newcommand{\OnetoOne}{\textcolor{oneToOneColor}{\textbf{1-1}}}
\newcommand{\OnetoN}{\textcolor{oneToNColor}{\textbf{1-n}}}
\newcommand{\N}{\textcolor{nColor}{\textbf{n}}}
\newcommand{\NtoN}{\textcolor{nToMColor}{\textbf{n-m}}}

\newcommand{\EVEA}{\textsc{ev-ea}}
\newcommand{\EVIA}{\textsc{ev-ia}}
\newcommand{\IVEA}{\textsc{iv-ea}}
\newcommand{\IVIA}{\textsc{iv-ia}}

\definecolor{lightpink}{RGB}{237,157,202}
\definecolor{lightred}{RGB}{210,121,121}
\definecolor{lightorange}{RGB}{230,170,50}
\definecolor{lightgold}{RGB}{210,194,121}
\definecolor{lightgreen}{RGB}{121,210,121}
\definecolor{lightaqua}{RGB}{121,206,210}
\definecolor{lightblue}{RGB}{121,124,210}
\definecolor{lightpurple}{RGB}{153,102,255}
\definecolor{red}{RGB}{178,34,34}
\definecolor{gray}{RGB}{166,166,166}

\newcommand{\eat}[1]{}

\newcommand{\teaserFigure}{
\teaser{
    \centering
    \includegraphics[width=\linewidth]{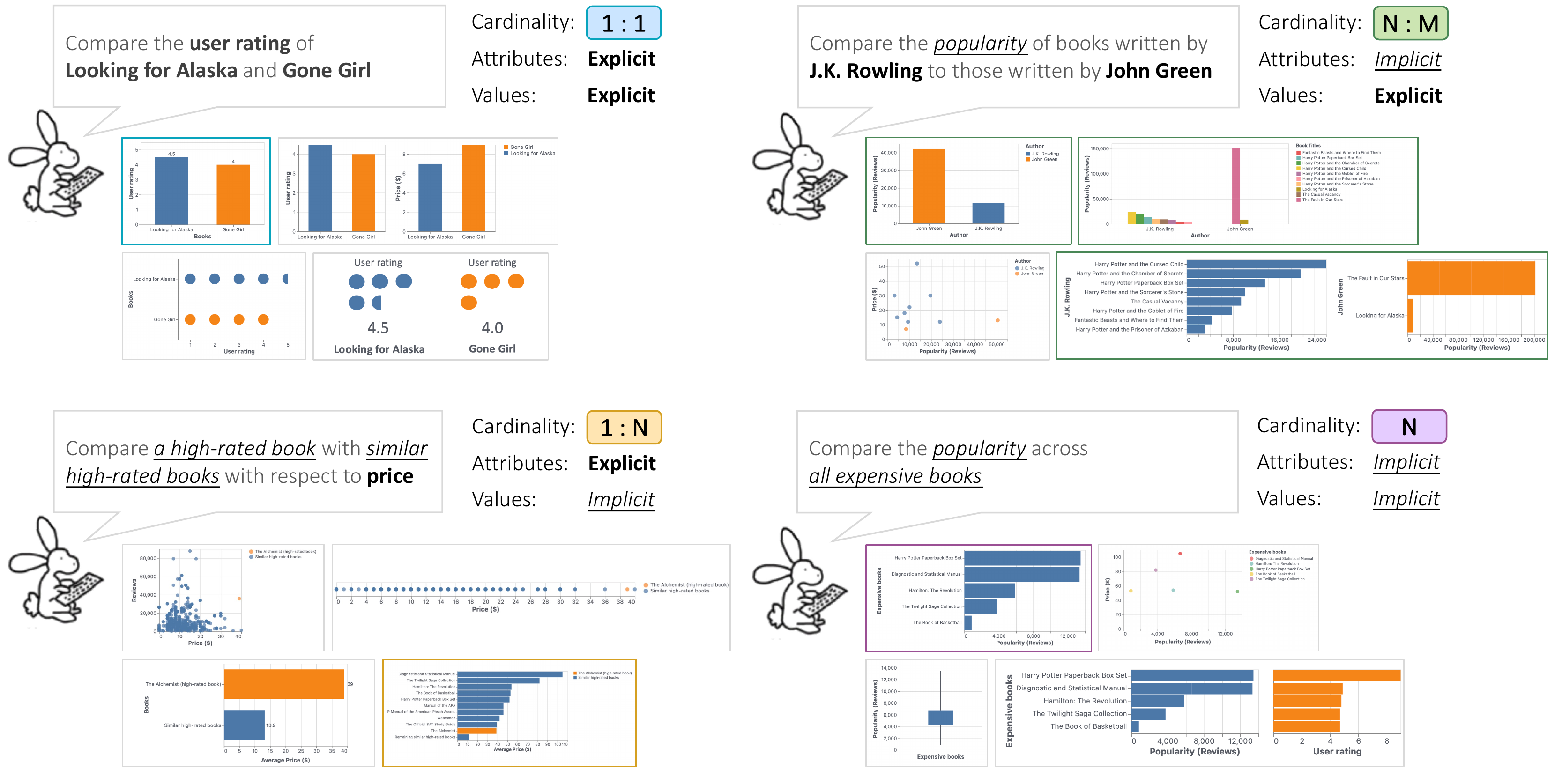}
    \caption{Four comparison utterances from our design space with varying cardinalities for the comparison entities (\OnetoOne, \OnetoN, \NtoN, \N) and different levels of concreteness (\textbf{explicit} and \emph{\underline{implicit}}). Each of these comparison utterances was included in our online survey in which participants ranked their preference for the different visualization types; the most preferred visualization(s) have a colored border.}
    \label{fig:wholeds}
}
}

\teaserFigure

\vgtcinsertpkg

\begin{document}


\firstsection{Introduction}
\maketitle

\begin{figure}[h!]
\centering
 \includegraphics[width = 0.8\columnwidth]{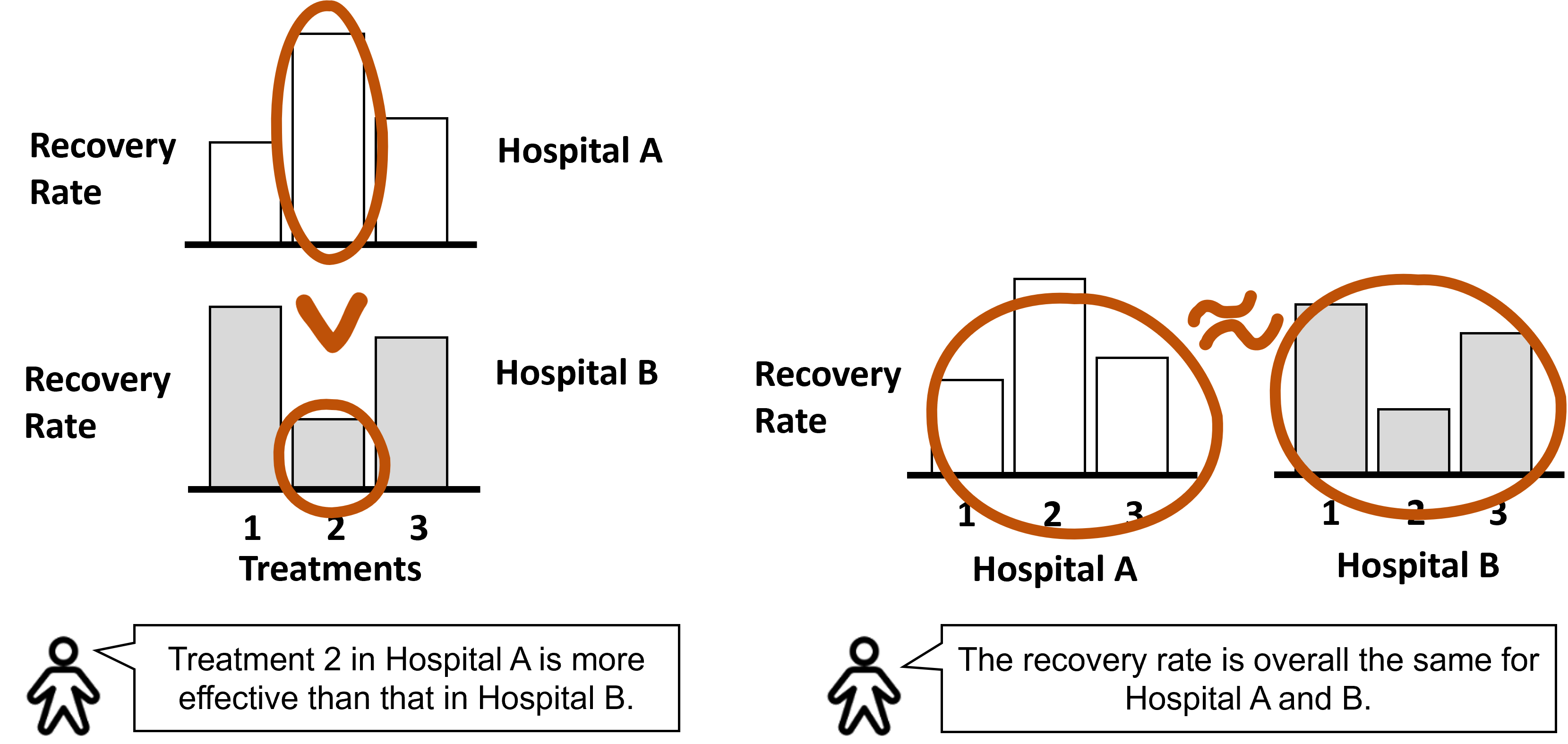}
 \caption{Examples of two sets of spatial arrangements for visual comparisons based on findings from Xiong et al.~\cite{xiong2021visual}. Results found that stacking small multiples vertically tends to elicit 1-to-1 comparisons of individual values in a set, whereas aligning them horizontally tends to elicit a comparison of the two sets as two groups.}
 \label{IntroComparison}
%  \Description{The figure shows 2 bar charts positioned in different ways. On the left they are positioned vertically and on the right they are positioned next to each other. Out of the 2 bar charts, the first bar chart shows 3 bars where the middle bar is the tallest showing the recovery rate for hospital A. The second bar chart also shows 3 bars where the first bar is the tallest, followed by the third bar, and then the middle bar showing recovery rate for hospital B.}
\vspace{-1em}
\end{figure}

Visual comparisons are a common and critical task in analytic workflows. When people read a visualization, each comparison they make can be thought of as an analytical sentence that describes common patterns, differences, and trends in the data \cite{shah2011bar}, such as \emph{``the recovery rate is overall the same for Hospitals A and B.''} Even seemingly-small design choices in a visualization can nudge viewers to see different patterns and produce different sentences \cite{burns2020evaluate, xiong2021visual}, further highlighting the complexities and nuances of comparison tasks. For example, Figure~\ref{IntroComparison} shows how the juxtaposition of small multiples, either vertically or horizontally, can elicit different comparisons across the two groups.

Recommendation systems attempt to provide smart defaults to help users gain insights into their data but focus only on common operations such as filtering and sorting. They do not provide recommendations for facilitating visual comparisons~\cite{eviza}. Recently, visual analysis interfaces have introduced natural language (NL) queries into recommendation systems~\cite{askdata} that enable users to type NL utterances and receive appropriate visualization responses. Such systems make extracting insights from data more intuitive, as the process of generating utterances exposes the mental model of how people think about comparisons~\cite{clark1972process, 47a7267cdf624c35831eb0dc765b434e, gleitman2007give, roth2012asymmetric}. However, the complexity of comparison expressions makes supporting NL-based visual comparisons a non-trivial task.

We identify two main challenges for building an effective recommendation system that can help users more easily make comparisons in visualized data. First, designing a visualization involves a series of decisions, each of which can afford different viewer takeaways \cite{xiong2021perceptual, xiong2021visual}. For example, past work has shown that people tend to make magnitude comparisons when viewing bar charts and trend comparisons when viewing line charts \cite{zacks1999bars, szafir2018good}. However, to the best of our knowledge, prior work does not propose a comprehensive mapping between comparison intents and the best visualization for the intended comparisons. 

Second, there exists semantic ambiguity and complexity in NL that makes the intent difficult to parse. These characteristics hinder the effectiveness of incorporating NL approaches to support visual comparison tasks~\cite{setlur:iui}. While \textbf{explicit} comparisons such as \emph{``compare the number of trees planted in location A to the number in location B''} are easy to decode, \underline{implicit} comparisons are difficult for visualization systems to interpret precisely. For example, consider the example, \emph{``Are the number of trees planted in location A tall?"}.
For comparison utterances that use vague modifiers such as ``tall'' \cite{hearst2019toward}, it is difficult to determine how the user intended for ``tall'' to be interpreted, i.e., what constitutes something being \emph{tall}?  For comparisons that contain underspecified references to the dataset, such as referring to ``performance'' when ``performance'' is not a variable explicitly contained in the dataset, it is similarly difficult to determine how a system should behave. Figuring out how to decode comparison intents from implicit utterances and underspecified references in NL for visualization use cases still remains an open question.

\pheading{Contributions}:
We contribute a preliminary design space of NL comparison utterances covering four cardinalities of comparisons and addressing ambiguities by differentiating implicit and explicit ways of expressing these comparisons. Through a series of interviews with 16 visualization novices and experts, we create a potential mapping between comparison utterances in our design space and different visualization techniques. We then empirically validate this mapping through an online survey with 77 participants to highlight three design implications for interpreting comparison utterances for visualization use cases. These design implications can inform the design of future NL-based recommendation systems to better support visual comparisons.
\section{Related Work}
Designing effective visualizations for comparison utterances is an ongoing area of research and can be categorized into three main themes: (1) comparisons in visualization, (2) comparisons in computational linguistics, and (3) natural language interfaces (NLIs) for visual analysis.

\subsection{Comparisons in Visualization}
Representing comparisons in data visualizations is an important aspect of a user's analytical workflow
and prior work has surveyed a variety of visualization solutions to better support comparisons. For example, Tufte discussed small multiples as a way to use the same graphic to display different slices of a data set for comparison~\cite{tufte_envisioning_1990}. Graham and Kennedy~\cite{grahamkennedy2007} surveyed a range of visual mechanisms to compare trees, while other surveys consider methods for comparing flow fields~\cite{Post1995}. Gleicher et al.~\cite{Gleicher2011} presented a broad survey with over 100 different comparative visualization tools from information visualization domains, organized by their comparative visual designs into a general taxonomy of visual designs for comparisons. Designs were grouped into three categories: juxtaposition, superposition, and explicit encodings.

The perceptual and cognitive science communities have also considered the problem of visual comparison for decades, including the issues around change blindness~\cite{rensink2002}. Other studies focused on comparing methods for specific tasks, such as the evaluation of scalar field comparisons~\cite{livingston2011} or brain connectivity graphs~\cite{alper2013}. Franconeri~\cite{47a7267cdf624c35831eb0dc765b434e, article} discussed several limitations in the mechanisms of perception that may have a direct impact on the design of comparison methods. For example, translations of an object are easy to compare, but texture, orientation, scale, space, and time may complicate comparison tasks~\cite{Larsen1978SizeSI,larsen:1998}.  

Existing work has largely focused on how \textit{visualization type} can influence viewer perception and decisions~\cite{zacks1999bars,xiong2019illusion,lee2016vlat}. For example, bar charts plot the data as discrete objects, motivating people to compare them as two distinct units (e.g., A is larger than B), while line charts plot data as one single object, eliciting the interpretation of trends, changes over time or relations (e.g., X fluctuates up and down as time passes)\cite{zacks1999bars}. Showing difference benchmarks on bar charts can not only facilitate a wider range of comparison tasks \cite{srinivasan2018s} but also increase the speed and accuracy of the comparisons \cite{nothelfer2019measures}. Charts that show probabilistic outcomes as discrete objects, such as a beeswarm chart, can promote a better understanding of uncertainties \cite{kay2016ish, hawley2008impact, Tait2010, garcia2009communicating}.

Comprehension of visual comparisons is an important aspect of determining their efficacy. Researchers have referenced knowledge from basic human perception research to generate design guidelines regarding graphical elements, such as the visualization color and shape. For example, in multi-class scatterplots, viewers can compare classes of scatterplots more effectively when they are colored differently compared to when they are plotted with different shapes \cite{duncan1989visual, gleicher2013perception}. Shah and Freedman~\cite{shah2011bar} investigated the effect of format (line vs. bar), viewers' familiarity, and their graphical literacy skills on the comprehension of multivariate data presented in graphs. The differences between people's perceptions of bar graphs and line graphs can be explained by differences in the visual chunks formed by the graphs based on Gestalt principles of proximity, similarity, and good continuity. Jardine et al.~\cite{JardineProxies} conducted an empirical evaluation on two comparison tasks: identify the ``biggest mean'' and ``biggest range'' between two sets of values. Their work showed that visual comparisons of these tasks are most supported by vertically stacked chart arrangements, and this pattern is substantially different across different types of tasks. 

 Much of this prior work has focused on exploring effective \emph{visual representation} of comparisons in data. However, with the prevalence of NLIs that support visual analysis, there is a need for such systems to be able to interpret comparison utterances and provide useful visualization responses to the user. Our work specifically focuses on better understanding the design space of mapping data comparisons expressed through \emph{language} with useful visual representations.

\subsection{Comparisons in Computational Linguistics}
The syntax and semantics of comparatives have been the topic of research in computational linguistics for quite some time. Comparative expressions which establish orderings among objects according to the amount or degree to which they possess some property is a basic component of human cognition~\cite{kennedy:2004}. Language constructs often provide concepts to express gradable concepts such as the explicit orderings between two objects (e.g., \emph{``the price of housing in the Bay Area is higher than in Texas''})~\cite{Sapir1944GradingAS}. Extensive prior work in computational linguistics has explored the semantics of comparison based on gradable concepts such as ``more'', ``less'', ``-er''~\cite{CRESSWELL1976261,Hamann2005COMPARINGST,Bierwisch1989TheSO,Klein1980-KLEASF,kennedy:1997,Schwarzchild2002QuantifiersIC,bakhshandeh-allen-2015-semantic}. 

The notion of vagueness in comparative language has previously been studied in the computational linguistics community~\cite{BacskaiAtkari2014TheSO,friedman-1989-general}. The semantics of comparatives can be vague as their interpretation depends on the context and the boundaries defining the comparative. For example, \emph{``when is it the safest time to fly?''} implicitly compares flying safety across different time periods, with ``safe'' being a fuzzy, subjective concept. Prior work has focused on the conceptualization and representation of vague knowledge. For example, Kessler and Kuhn presented a corpus of annotated comparison sentences from English camera product reviews. Each sentence contains the comparative predicate that expresses the comparison, the comparison type, the two entities that are being compared, and the aspect in which they are compared~\cite{Kessler2014}. 

While linguistic vagueness has been explored for comparative expressions along with their semantic variability, little work has been done in determining how best to \emph{visually} represent comparatives based on these variations, especially in the context of visual analysis.

\subsection{Natural Language Interfaces (NLIs) for Visual Analysis}
Much of the effort for supporting NLIs for visual analysis~\cite{datatone,eviza, dhamdhere2017, askdata,thoughtspot,ibmwatson,powerbi} focuses on interpreting a user's analytical intent by providing useful visualization responses. The methods of interpreting intent typically rely on explicitly named data attributes, values, and chart types in the user's input queries.  Ask Data~\cite{setlur:iui} supports analytical expressions in NL such as grouping of attributes, aggregations, filters, and sorts. The system also handles impreciseness around vague numerical concepts such as ``cheap'' and ``high'' by inferring a range based on the underlying statistical properties of the data. Hearst et al.~\cite{hearst2019toward} explore appropriate visualization responses to vagueness by interpreting singular and plural superlatives (e.g., ``highest price'' and ``highest prices'') and numerical graded adjectives (e.g., ``higher'') based on the shape of the data distributions. Law et al.~\cite{Law2020CausalPI} investigated how the visual design of answers
to \emph{why} questions might influence user perceptions of a question-answering system. They found that users have a strong tendency to
associate correlation with causation when systems do not provide clear explanations for the answers.

The space of analytical expression in NL is rich and much more nuanced than what these interfaces currently support. A study was conducted to assess NL input to visualization
systems~\cite{setlur:iui} where $75$ participants were asked to write NL queries based on five underlying datasets (i.e., bird strikes, world indicators,
superstore, mutual funds, and Olympic medals). Of the $578$ NL queries, common ones included \emph{``Are there more strikes on takeoff or landing?''}, \emph{``Are certain seasons more dangerous?''}, and \emph{``Which country has more female medalists?''} The data suggests that when participants were
not restricted in the format of expression, they often chose to specify utterances with an underlying intent to either explicitly or implicitly compare values. Current NL systems, however, do not explore how utterances about comparisons ought to be interpreted even though such forms of intent are prevalent. 

Our paper identifies a gap in mapping how users express comparative utterances in NLIs to appropriate visualizations. To address this problem, we explore a design space connecting language and visual representation for a range of comparison utterances, varying in what is being compared and how these comparisons are specified. Finding from our work provide implications for NLIs and recommendation tools to better interpret and support comparisons.
\section{Design Space of Comparison Utterances}
In visual analytics, Gleicher et al. \cite{Gleicher2018ConsiderationsFV} define comparisons as an analytical task involving two components: a set of targets, i.e.,~the set of items being compared, and an action performed on the relationships among these targets, e.g., similarities and differences. We extend this definition of comparison as the task of identifying similarities and differences between two or more categories of data \textbf{values} based on one or more data \textbf{attributes} shared among those categories. For the remainder of this paper, we refer to this extended definition of comparisons.

In this work, we explore the intricacies of language constructs used to express comparisons and particularly focus on understanding the semantic variations in comparison expressions and how those variations influence the corresponding visual representations. To this end, we define a \emph{comparison utterance} as a textual sentence, inputted by a user to an NLI, for example, that expresses the intent of performing a comparison with a given dataset. An example of a comparison utterance is \emph{``compare the sales for Washington and California,''} where ``Washington'' and ``California'' are the values, and \texttt{sales} is the attribute.  Inspired by the language constructs from the computational linguistics literature~\cite{shapiro:2006,keefe:2008}, we describe our design space for comparison expressions in terms of the \textbf{cardinality} and \textbf{concreteness} of the comparisons.

%Comparison utterances comprise various semantic variations for the comparatives and gradable predicates expressed through language~\cite{}. Specifically, we consider two prevalent parameters of variation identified in computational linguistics literature~\cite{}:

\pheading{Cardinality}: Comparison utterances can express relationships between individual entities, e.g., \myquote{compare the effectiveness of treatment A to treatment B}, as well as relationships between individuals and larger sets containing multiple entities, e.g., \myquote{compare the effectiveness across all treatments}. Similar to the process described by Xiong et al.~\cite{xiong2021visual}, we list four cardinalities for the entities in comparison utterances in the context of visual analysis:

\pheading{\OnetoOne}: Compare one entity to another entity\\
e.g., \emph{``compare the IMDB ratings of Squid Game and Midnight Mass''}

\pheading{\OnetoN}: Compare one entity to another set of multiple entities\\
e.g., \emph{``compare the performance of Starling to other PG-13 movies''}
%\emph{``compare the number of times Squid Game was watched to other thriller TV shows''}
%\emph{``compare the amount of times Squid Game was watched as opposed to all other thriller TV shows''}

\pheading{\N}: Compare multiple entities within the same set\\
e.g., \emph{``compare the budgets across all US movies''}

\pheading{\NtoN}: Compare one set of entities to another set\\
e.g., \noindent\emph{``compare crime shows to thriller shows in terms of box office''}

\pheading{Concreteness}: Comparison utterances can \emph{explicitly} mention data attributes and values to be compared; for example, the comparison expression \myquote{compare the number of silver medals won by Rebecca Adlington to all other participants in the Women's Swimming Event} explicitly refers to the values to consider (``Rebecca Adlington'' and ``all other participants in the Women's Swimming Event'') as well as the exact data attribute to use in the comparison (\texttt{silver medals}). 
%\vidya{add an example for an explicit comparison utterance}. 
Due to language variations, we account for plurality (e.g.,~``book'' and ``books''), spelling variation (e.g.,~``color'' and ``colour''), and stemming (e.g.,~``costly'' and ``cost'') for explicit references.

Comparison utterances can also be underspecified when one or more data attributes or values are \emph{implicitly} referenced in the comparison~\cite{sentifiers}. For example, the utterance \myquote{compare the \textbf{popularity} of all movies in 2021} could be considered implicit if the dataset does not contain an attribute or value that explicitly matches the token ``popularity.'' Rather, other attributes such as \texttt{number of reviews} or \texttt{user rating} may be more appropriate to consider when comparing the popularity of the movies. In addition, implicit comparison utterances can employ language constructs such as gradable vague modifiers like ``low'', ``high'', or ``cheap'' when comparing data, e.g.,~\myquote{compare athletes who won a \textbf{high} number of gold medals}. These gradable adjectives are often mapped to an upper or lower range of data values in a data column~\cite{setlur:iui,hearst2019toward}.

\subsection{Identifying the design scope of comparisons}
Both the cardinality and concreteness of the comparison utterance can impact the desired visual characteristics to perform the corresponding comparison task.
Within each cardinality, there are theoretically 16 combinations of comparisons that we can identify via a truth table by permutating whether each of the two data values and data attributes are implicit or explicit ($2\times2\times2\times2)$.
%as shown in table [todo]. % For example, a data value can be explicit while the data attribute can be explicit such as ``compare the box office numbers of all high rated movies made is the US`` where box office numbers is an explicit data attribute and high rated movies is an example of implicit data vales and this utterance represents the n cardinality. 
However, not every combination produces a valid or natural comparison scenario. 
For example, the order of elements for degrees of concreteness does not fundamentally change the resulting comparison task; in other words, the comparison \myquote{compare the performance of Starling to other high rated movies} (an explicit value followed by an implicit value) is essentially the same as the comparison \myquote{compare the performance of high rated movies to Starling} (an implicit value followed by an explicit value). We can therefore reduce the space to focus on utterances that are order agnostic. Furthermore, while we initially consider a design space of two values with two corresponding attributes, comparisons that involve \emph{different} attributes for each value are rather nonsensical like comparing ``apples'' and ``oranges'', e.g.,~\myquote{compare the gold medals obtained by Rebecca Adlington to the bronze medals obtained by Nathan Ghar-Jun Adrian}. Finally, we chose to exclude comparisons with different degrees of concreteness for the values as many of these comparisons felt artificial compared to real-world comparison utterances, e.g., \myquote{compare a low budget TV show to Squid Game with respect to their popularity}.

The final four combinations of concreteness in our design space are as follows: \textit{explicit} data values paired with \textit{explicit} data attributes (\EVEA); \textit{explicit} values and \textit{implicit} attributes (\EVIA); \textit{implicit} values and \textit{explicit} attributes (\IVEA); and \textit{implicit} values paired with \textit{implicit} attributes (\IVIA).
Our final design space, therefore, includes a total of 16 utterances across four cardinalities (\OnetoOne, \OnetoN, \N, \NtoN) and four combinations of concreteness (\EVEA, \EVIA, \IVEA, \IVIA).
The final design space along with example queries for both datasets can be found in our supplemental material. 
\footnote{
\label{footnote:supplemental} Additional details regarding the experimental protocol, datasets, and all the comparison utterances can be found in \href{https://osf.io/zp6kh/?view_only=dfcf6012d3654fd99884ae8523e4f1aa}{supplemental material}.}

\section{Study 1: Designing Visualizations for Comparisons}
\label{sec:study1}

We conducted user studies with visualization experts and non-experts to better understand how they visualize data for comparison utterances.

\subsection{Participants}
We recruited 16 participants (six female, ten male). All participants were fluent in English and self-reported being comfortable designing visualizations. Eight participants (four female, four male) self-reported as data visualization experts with an average of $\sim$11 years of visual analytics experience and were recruited from a visual analytics company. The data visualization experts had a variety of job backgrounds including three data visualization consultants, two user experience designers, one business strategist, one visualization lead expert, and one solution engineer. The other eight (two female, six male) self-reported as data visualization non-experts and were recruited from an academic institution. All of the non-expert participants had a computer science or engineering background, with six being graduate students and two being undergraduates. Participants were recruited via a company mailing list, Slack, and using snowball sampling \cite{Gelo2008QuantitativeAQ}. The students were offered to enter a \$30 gift card raffle as an incentive to participate, while the participants recruited from the visual analytics company were not offered any financial incentive due to company policy. 

\subsection{Procedure}
Each study session lasted 60 minutes and was conducted remotely via Zoom. Two researchers supported each session: one facilitator and one note-taker.
All sessions were video recorded. Field notes were expanded to a video log after the study through partial transcription of the videos. The video log (and raw video for reference) was then qualitatively coded for high-level themes. 

Participants recruited from the academic institution were asked to fill out a consent form to indicate their interest in receiving a gift card. All participants were asked for their consent for the session to be recorded, and participants were then asked to share their screens in order to make the session more interactive between the participant and the facilitator. 

During each session, participants were first introduced to the study and asked about their visualization background and current role. Participants were then given instructions for the study and were asked to draw one or more visualizations for four comparison queries.
The main study prompt was as follows:
\emph{``Imagine that you are a visualization recommendation tool that generates visualization(s) when prompted by a user. Assume that the user imported a [Netflix or Olympics] dataset, and inputted some utterances describing what they want to see from the data. Your task is to generate a visualization or multiple visualizations (if you think that it can be visualized in multiple ways) as a response to the utterance and present it back to the user. Please sketch your visualization response that you think will best respond to their request.''}

Participants were first provided a weblink to a Google Sheet containing the dataset along with a metadata summary sheet. They then read four comparison utterances one at a time and used Google Jamboard~\cite{jamboard} to sketch out the visualizations. Participants were encouraged to think aloud during the study, and we employed a question-asking protocol to elicit qualitative feedback. Responses to common questions were documented ahead of time so that the facilitator could provide consistent responses to every participant. 
% We had a lot of participants ask questions when they encountered implicit data values or data attributes such as, \textit{What is performance?} To which our response was, \textit{What do you think performance means?}. Other question was, Can I make changes in the dataset?" to which our response was, "Yes, of course." and another question, "Can I make this sketch interactive?" to which we had multiple responses such as "What do you mean by interactive?", "Why do you think it should be interactive?", or "How do you think interactivity helps?". 
% Additional details regarding the experimental protocol, datasets and their metadata summaries, and all the comparison utterances can be found in supplemental material at \cite{supplemental}.

% We asked interviewees to sketch how they would visually represent responses to various comparison utterances. 
% We asked open-ended questions and encouraged interviewees to describe what they want to see from the data, such as "how would you design it differently" or "why do you think it should be represented in so many ways".  We probed them to sketch multiple visualizations for each utterance if possible and also reason about which they would prefer.

\subsection{Experimental Design}
\begin{figure}[!tp]
    \centering
    \includegraphics[width=\columnwidth]{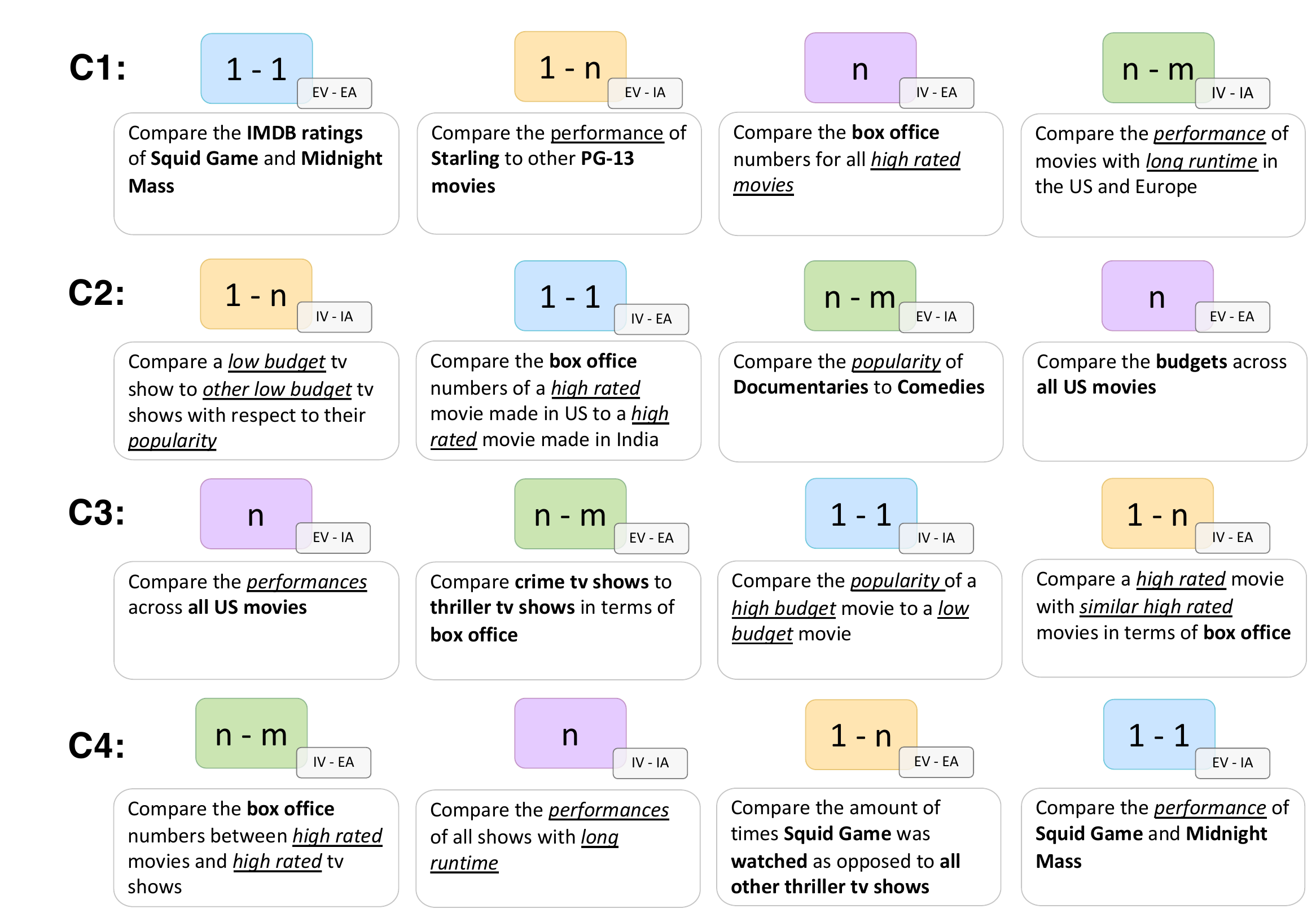}
    \vspace{-20px}
    \caption{A Graeco-Latin square with example utterances from the Netflix dataset and four participants came across each condition.}
    \vspace{-5px}
    \label{fig:graecosquare}
    % \Description{This figure shows how each condition was formed using the Graeco latin square with example utterances. The details of this figure are explained in the text.}
\end{figure}

This study employed a $4\times4$ Graeco-Latin square design that covered four different cardinalities (\OnetoOne, \OnetoN, \N, and \NtoN) and two levels of concreteness (I:~implicit and E:~explicit) for the data values~(V) and data attributes~(A). Each participant came across all four cardinalities and all four combinations of concreteness (\EVEA, \EVIA, \IVEA, \IVIA); the $4\times4$ Graeco-Latin square design balances the order in which the participant comes across the utterances. %Thus, there are four conditions where each participant was assigned to only one condition.
The conditions are shown in Figure \ref{fig:graecosquare} and described as follows:
\begin{tight_itemize}
\item \textbf{Condition 1}: \OnetoOne\ \EVEA, \OnetoN\ \EVIA, \N\ \IVEA, \NtoN\ \IVIA 
\item \textbf{Condition 2}: \OnetoN\ \IVIA, \OnetoOne\ \IVEA, \NtoN\ \EVIA, \N\ \EVEA
\item \textbf{Condition 3}: \N\ \EVIA, \NtoN\ \EVEA, \OnetoOne\ \IVIA, \OnetoN\ \IVEA
\item \textbf{Condition 4}: \NtoN\ \IVEA, \N\ \IVIA, \OnetoN\ \EVEA, \OnetoOne\ \EVIA
\end{tight_itemize}

% Participants were assigned to one of two datasets (described in the following section) and one of these conditions. Therefore, for each dataset, there were two different participants who encountered that condition and eight participants in each cardinality type.

\subsection{Experimental Dataset}

In each condition, participants interacted with one of two datasets: a Netflix dataset~\cite{bansal_2021} and an Olympics medals dataset~\cite{rgriffin_2018}. The Netflix dataset contains 13 attributes including seven categorical (e.g., Title, Genre), six numeric (e.g., IMDB Rating, Box office), and one temporal (Release year). The Olympics dataset also contains 13 attributes composed of six categorical (e.g., City, Event), six numeric (e.g., Height, Gold), and one temporal attribute (Year). We chose these datasets as they contain a combination of continuous and categorical variables that can support a variety of comparison utterances. The datasets were also familiar to a wide set of participants and hence could be easily interpreted. We used a subset of 60-70 rows for each of the two datasets to prevent participants from becoming overwhelmed with the amount of data when performing the tasks.

\subsection{Pilot Studies}
We conducted nine pilot studies to test the study protocol and stimuli. Given the remote experimental setup using a combination of online tools, the goal of the pilots was to streamline the user study process and resolve any points of confusion, particularly around the instructions and utterances. We conducted, transcribed, and analyzed $\sim$540 minutes of pilot interviews before conducting the final user study.

Based on feedback from the pilot participants, we updated the instructions for the study and refined the utterances and datasets. To better orient the participant about the dataset and its attributes and values, we included a metadata summary. We also incorporated a short tutorial in the study protocol to familiarize participants with the various drawing options in Jamboard and data manipulation features in Google Sheets.  %We also observed several pilot participants wanting to manipulate the data based on the comparison utterances. We therefore included a brief tutorial in the study to get participants acquainted with Google Sheet features such as filter and sort.
% \arjun{This whole section can be removed and replaced with one sentence somewhere at the start of 4.2: ``The study protocol was iteratively designed through nine pilot studies."}

\subsection{Study 1: Qualitative Analysis}

\noindent In total, we gathered $\sim$960 minutes of video and audio recordings, transcribed them using Amazon AWS Transcribe \cite{kranz_1986}, and then placed each unique statement into our analysis software, a shared spreadsheet. We extracted the following details from the recordings: 

\pheading{Visualization design:} We extracted details about the visualization \emph{type} (e.g., bar chart, scatter plot) and \emph{encoding} (e.g., color, size), which included details about the mapping between data values and visual properties. We also considered high-level properties of the visualization layout such as the \emph{orientation} (e.g., vertical or horizontal), \emph{arrangement} ~\cite{xiong2021visual} (e.g., adjacent, overlaid), and whether or not the visualization used a \emph{small multiples} design~\cite{5473226}.

\pheading{Visualization details:} We also extracted additional narrative characteristics of the visualization such as the use of \emph{highlighting} to emphasize certain features and the use of \emph{annotations} (e.g., labels, reference lines, legends). We also recorded whether or not the visualization was \emph{sorted}.

\pheading{Visualization or data transformations:} An important characteristic of this study was how participants \emph{interpreted} implicit data values and attributes; if participants identified multiple interpretations, we also recorded which of them the participant preferred. As part of the design process, we recorded how the data was \emph{filtered} before drawing; for instance, for the utterance \myquote{compare the performance of Starling with other PG-13 movies,} some participants chose only the ``top 3'' movies to compare with Starling, whereas others chose 5, and so on. Finally, we extracted the desired \emph{interactive} features of the visualization described by participants, such as the inclusion of a tooltip or data label on hover, or other ways in which the visualization could change on demand.

\pheading{Participant details \& preferences:} In addition to \emph{demographic} characteristics like the participant's occupation, and expertise, we extracted the \emph{number} of visualizations created during the study and in cases where multiple visualizations were drawn for an utterance, we also extracted which visualizations the participant \emph{preferred} and why.\\

\subsubsection{Interpreting ambiguity of implicit values and attributes}
One of the main goals of this study was to understand how participants interpret the concreteness of the data values and data attributes. Participants came across multiple implicit data \textbf{values} such as ``high rated movie'', ``high budget movie'', ``tall athlete'', ``successful wrestler'', and ``high achieving'', as well as implicit data \textbf{attributes} such as ``performance,'' ``popularity,'' ``physique,'' ``achievements,'' and ``long runtime''. Additional details regarding the datasets and their metadata summaries can be found in the \href{https://osf.io/zp6kh/?view_only=dfcf6012d3654fd99884ae8523e4f1aa}{supplementary material}.

\pheading{Netflix Dataset:} The following list describes the implicit data values and data attributes, their implicit type (i.e., vague modifier or underspecification), and examples of how they were commonly interpreted by participants for the Netflix dataset:
\begin{tight_itemize}
    \item \textbf{performance} (\textit{vague modifier}): [Watched] $OR$ [IMDB rating] $OR$ [Rotten tomatoes rating]  $OR$ ROI/Profit  [Box office $-$ Budget] $OR$ [Box office]  
    \item \textbf{popularity} (\textit{vague modifier}): [Box office $-$ Budget] $OR$ [Watched]  
    \item \textbf{high rated} (\textit{underspecified}): [Rotten tomatoes rating] $> 80/100\ OR$ [Rotten tomatoes rating] $>AVG$(Rotten tomatoes rating$)\ OR$ [IMDB rating] $> AVG($IMDB rating$)\ OR$ [IMDB rating] $> 8/10$
    \item \textbf{high budget} (\textit{underspecified}): [Budget] $> AVG($Budget$)\ OR$ [Budget] $> 95^{th} percentile$
    \item \textbf{low budget} (\textit{underspecified}): [Budget] $< AVG($Budget$)\ \ OR$ [Budget] $< 5^{th} percentile\ OR$ lowest budget value
    \item \textbf{long runtime} (\textit{underspecified}): [Duration] $> 100$ minutes $OR$ [Duration] $>\ 80^{th} percentile$
\end{tight_itemize}

\pheading{Olympics Dataset:} The following list describes some of the implicit values and attributes, their implicit type, and examples of how they were interpreted for the Olympics dataset:

\begin{tight_itemize}
    \item \textbf{achievements} (\textit{vague modifier}): $SUM($Gold, Silver, Bronze$)$
    \item \textbf{performance} (\textit{vague modifier}): $SUM$(Gold, Silver, Bronze) $OR$ weighted metric of [Gold, Silver, Bronze]
    \item \textbf{physique} (\textit{vague modifier}): [Height] $OR$ [Weight] $OR$ metric of [Weight, Age, Height]
    \item \textbf{tall athlete} (\textit{underspecified}): [Height] $> AVG($Height$)\ OR$ [Height] $> MEDIAN($Height$)\ OR$ [Height] $\geq 180$cm
    \item \textbf{short athlete} (\textit{underspecified}): [Height] $< AVG($Height$)\ OR$ [Height] $< MEDIAN($Height$)\ OR$ [Height] $< 180$cm
    \item \textbf{strong physique} (\textit{underspecified}): [Weight] $> AVG($Weight$)\ OR$ [Weight] $>$ weighted sum of [Weight, Height, Age]
    \item \textbf{successful player} (\textit{underspecified}): $SUM($ Gold, Silver, Bronze $) > 0$
    \item \textbf{high achieving} (\textit{underspecified}): [Gold] $>$ 0
    \item \textbf{young athlete} (\textit{underspecified}): [Age] $< MEDIAN($Age$)$ $OR$ [Age] $<$ 20  
    \item \textbf{top-winning} (\textit{underspecified}): [Gold] $> 0\ OR$ weighted sum of [Gold, Silver, Bronze] $> 0$
    \item \textbf{senior athlete} (\textit{underspecified}): [Age] $> AVG($Age$)$
\end{tight_itemize}

% \noindent The implicit data values or data attributes were sometimes interpreted in a single context or multiple contexts from the list given above and according to the dataset.
\noindent While a few of the participants interpreted the implicit data values or attributes in only a single context from the list given above, most of them interpreted them in multiple contexts.
In the case of underspecified values, we notice a general trend of participants interpreting values as being above or below average. In the case of vague modifiers, participants usually picked all the attributes from the dataset related to the implicit value or attribute and mapped all of them to it. For example, P9 explained their approach to resolving the ambiguity as follows: \textit{``performance is kind of a vague thing. So let's just assume number of medals. That would be the best thing to assume. So performance is how many accolades they got.''}. Participants also tended to use some form of a weighted average, algorithmic, or parametric approach to interpret the implicit terms; for example, P4 interpreted ``high rated'' as \emph{``I would statistically identify the high ratings more of like in a statistical or algorithmic way.''}

\subsubsection{Developing clear charts for the intended comparison task}
Some participants found the decision process for developing the visualization designs particularly difficult when trying to handle aggregation or other forms of data transformation. Reflecting on their role as ``a visualization recommendation tool,'' one participant explained that \emph{``I would prefer for a recommendation engine to say don't look at the mean, look at the median. But if it's a normal distribution, I'd prefer it to say, let's look at the mean... you can trust us and that we are automatically providing the best results.''} [P4].

% When it comes to how best to support comparisons, the visualizations should leverage appropriate design decisions (especially when it comes to data transformations) to ensure that the chart provides the necessary information to complete the viewer's comparison task.
A few participants suggested leveraging appropriate design decisions, such as highlighting certain parts of the visualizations or adding data transformations, to make the comparison easier for a viewer to comprehend.
P9 shared insights on how design decisions can support comparison tasks: \emph{``So even something as simple as like a dot on the left to draw your eye to Rebecca because she is central to the question... or I would do the same chart, but put her at the very top, so she's floated to the top and then beneath her we're still in descending order.''}.

Another participant argued that the initial visualizations should be rather simple when describing their 2-bar bar chart: \emph{``I don't want to start with more than they asked. The two bars next to each other, I feel like they're good if there are not too many things that you're comparing. So if it's just two movies, I feel like this works really well.}'' [P4].\\

\subsection{Study 1: Quantitative Analysis}

% \arjun{I'm not the study expert so please feel free to ignore but: given that this was an elicitation study, do we need all the statistical info on time spent etc.? Can't we just summarize the number of visualizations (current 4.7.2) and report average time participants spent on the tasks?}

We analyzed the time spent designing visualizations based on the four cardinalities and level of concreteness, as well as the number of visualizations that the participants sketched. Additional details can be found in the supplementary materials.
\newline
\vspace{-2mm}

\noindent \textbf{Time Spent:} Two-way ANOVAs comparing the time people spent for each of the 16 queries from both datasets showed no difference between cardinalities, concreteness, nor datasets $(p > 0.05)$. On average, participants spent 9.75 min (SD = 0.52) per query.
\newline
\vspace{-2mm}

\noindent \textbf{Number of Visualizations:} Two-way ANOVA comparing the number of visualizations (including the scratched out ones) that participants sketched for each query in both datasets suggests that there is no significant effect of concreteness, cardinality, or their interaction. On average, participants drew 2.25 visualizations (SD = 0.15) per query.  
\newline
\vspace{-2mm}

% \newcolumntype{C}[1]{>{\centering\arraybackslash}m{#1}}
% \begin{table}[h]
%     % \begin{tabular}{|m{2cm}|m{7cm}|m{3cm}|}
%     \setlength\extrarowheight{2pt}
%     \centering
%     \begin{tabular}{|C{2cm}|C{2.5cm}|C{2.5cm}|}
%     \hline
%         \textbf{Cardinality} & \textbf{\% of visualizations with interaction} & \textbf{\% of visualizations with annotation} \\ \hline
%         \OnetoOne & 23.3 & 36.7 \\ \hline
%         \OnetoN & 25.7 & 42.9 \\ \hline
%         \N & 45.2 & 45.2 \\ \hline
%         \NtoN & 24.4 & 43.9 \\ \hline
%         \textbf{Concreteness} & & \\ \hline		
%         \EVEA & 13.2 & 34.2 \\ \hline
%         \EVIA & 28.6 & 40.5 \\ \hline
%         \IVEA & 37.0 & 51.9 \\ \hline
%         \IVIA & 36.1 & 38.9 \\ \hline
%     \end{tabular}
%     \vspace{4px}
%     \caption{The percentage of visualizations that participants described as \emph{interactive} or \emph{annotated} for each cardinality and concreteness.\jane{I wonder if this table could get smaller somehow?}}
%     \label{tab:interact-annotate}
% \end{table}

\newcolumntype{C}[1]{>{\centering\arraybackslash}m{#1}}
\begin{table}[t!]
    % \begin{tabular}{|m{2cm}|m{7cm}|m{3cm}|}
    \setlength\extrarowheight{2pt}
    \centering
    \caption{The percentage of visualizations that participants described as \emph{interactive} or \emph{annotated} for each cardinality and concreteness.}
    \begin{tabular}{|C{0.48cm}|C{0.48cm}|C{0.48cm}|C{0.55cm}|C{0.83cm}|C{0.83cm}|C{0.83cm}|C{0.83cm}|}
        \multicolumn{8}{c}{\textbf{Percent of visualizations with interaction}} \\ \hline
        \multicolumn{4}{|c|}{\textbf{Cardinality}} & \multicolumn{4}{c|}{\textbf{Concreteness}} \\ \hline
        \OnetoOne & \OnetoN & \N & \NtoN & \EVEA & \EVIA & \IVEA & \IVIA \\ \hline
        23.3 & 25.7 & 45.2 & 24.4 & 13.2 & 28.6 & 37.0 & 36.1 \\ \hline
        \multicolumn{8}{c}{\textbf{}} \\
        \multicolumn{8}{c}{\textbf{Percent of visualizations with annotation}} \\ \hline
        \multicolumn{4}{|c|}{\textbf{Cardinality}} & \multicolumn{4}{c|}{\textbf{Concreteness}} \\ \hline
        \OnetoOne & \OnetoN & \N & \NtoN & \EVEA & \EVIA & \IVEA & \IVIA \\ \hline
        36.7 & 42.9 & 45.2 & 43.9 & 34.2 & 40.5 & 51.9 & 38.9 \\ \hline
    \end{tabular}
    \vspace{4px}
    \vspace{-20px}
    \label{tab:interact-annotate}
\end{table}

We also recorded the number of visualizations with \emph{interactive} features and \emph{annotations} for each dimension (Table \ref{tab:interact-annotate}).
Based on the four cardinalities, we see that cardinality \N\ had the highest percentage of visualizations with interactive features, followed by cardinality \OnetoN. In terms of concreteness, \IVEA\ had the highest percentage of visualizations with interactive features, closely followed by \IVIA. When P11 was asked why they added interactivity to their sketch, they commented, \emph{``I don't know anyone who would really want to look at gold and silver, but I would look at gold medals versus all medals seems like a reasonable question... being able to do things like add events at years, filter by medals, they seem to be the adjacent questions to me in this.''}

Regarding the percentage of annotated visualizations, \N\ had the highest, followed by \NtoN\ and \OnetoN. In terms of concreteness, \IVEA\ had the highest percentage of annotated visualizations, followed by \EVIA, and then \IVIA. Many participants emphasized the importance of annotations and highlighting to make the comparisons clearer. P2 explained that \emph{``you could highlight a specific show... so it stands out and then all the other ones would be a different color.''} This strategy was common across participant sketches and informed the design of the representative visualizations in Study 2 (Section~\ref{sec:study2}).

\subsection{Identifying Common Visualizations per Cardinality}
\label{ExtractedVisualizations}
\begin{figure}[t]
    \centering
    \includegraphics[width=\columnwidth]{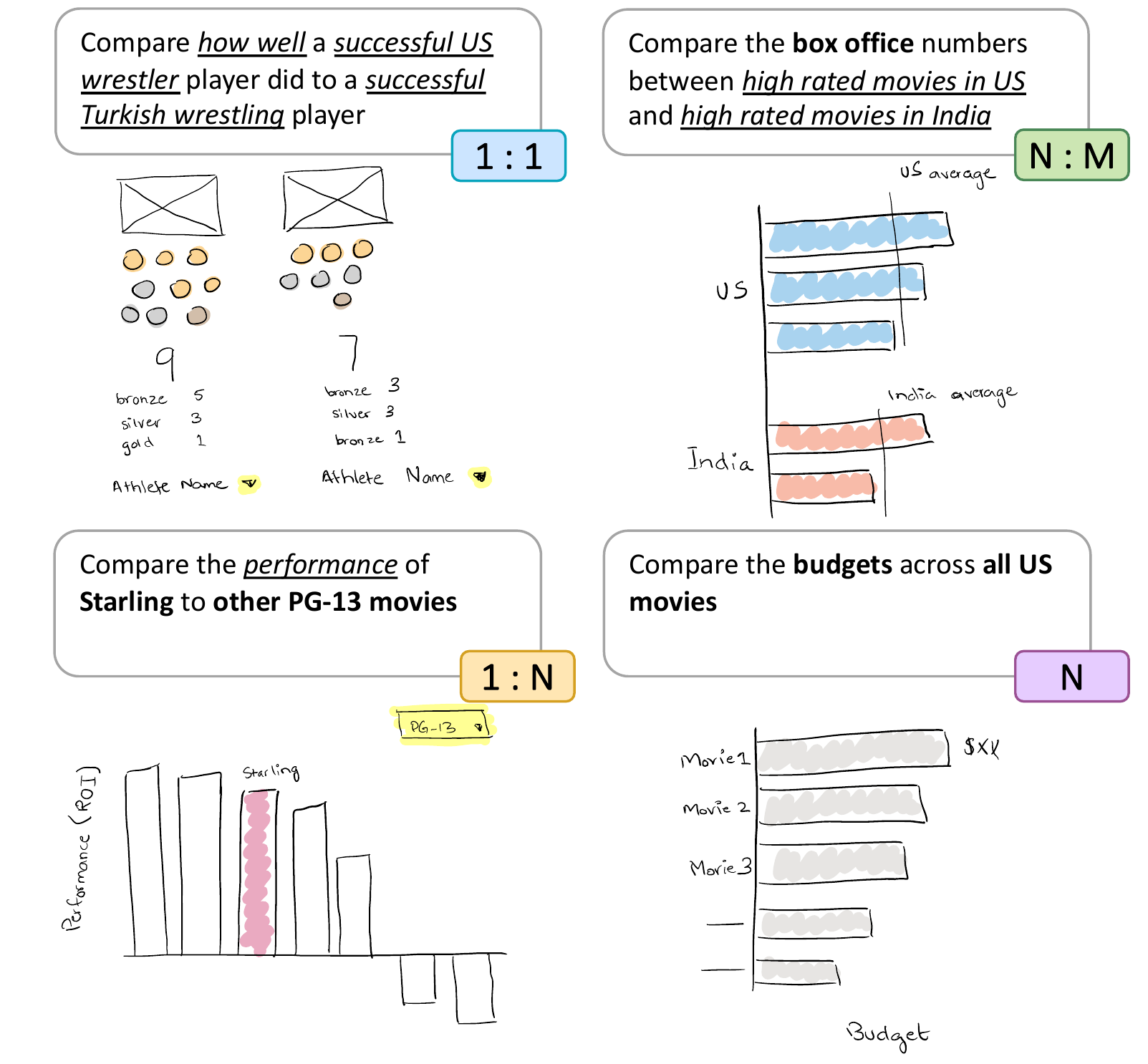}
    \vspace{-20px}
    \caption{Sketches elicited from Study 1. Charts are recreated for enhanced clarity based on the original sketches and interview script.}
    \vspace{-12px}    
    \label{fig:sketches}
    % \Description{The figure shows 4 visualizations for the 4 cardinalities with an example utterance for each. For 1 to 1, the utterance is; compare how well a successful US wrestler player did to a successful Turkish wrestler player. The vis shows 9 circles on the left and 7 circles on the right. Below the circles is the number 9 and 7 respectively. Below the number is individual count for gold, silver and bronze medals. Finally there is a players' name on the bottom with a dropdown option. For 1 to n, the utterance is; compare the performance of Starling to other PG-13 movies. The vis for it is a vertical bar chart sorted from left to right with the bar for Sterling colored in pink with performance (ROI) on the y-axis. For n, the utterance is; compare the budgets across all US movies. The vis for it is a horizontal bar chart sorted from top to bottom with budget on the x-axis and movies on the y-axis. The budget (number) annotated on the top of the bar. For n to m, the utterance is; compare the box office numbers between high rated movies in US and high rated movies in India. The vis for it is a horizontal bar chart divided in the middle. Top bars depict US movies and the bottom bars depict movies in India. There is a line that shows the average budget of movies in US and India respectively.}
\end{figure}

Based on the participants' sketches, we created illustrations of the visualizations using Notability software~\cite{notability}; a sample set shown in Figure~\ref{fig:sketches}. We clustered the visualizations based on the four cardinalities. Two authors went through the visual clusters to find commonalities between the charts. We initially thought that the top three visualizations from each cardinality would be sufficient to represent the common visual charts. However, after some extensive examination of the visual clusters, the top four visualizations for each cardinality were chosen. % as we thought this number best represents the common visualizations from each cardinality.

The top four visualizations were chosen based on the occurrence of the chart type, arrangement, and orientation of the chart. We then looked at whether and how these charts were annotated in order to include those annotations in our generalized drawings. We also looked in detail into how the implicit values and attributes were interpreted by the participants. We commonly noticed the introduction of interaction, grouped/small multiples/stacked, just mapping of one attribute or value from the dataset in visualizations that were sketched for levels of concreteness involving implicitness. As the focus of this paper is not to look into interactions in visualizations, we decided to map multiple or single attribute(s)/value(s) in the levels of concreteness that involve ambiguity. In the following paragraphs, we describe in detail the commonalities in the sketches by the participants and the process by which 16 visualizations were extracted, i.e., the top four for each cardinality.

% \arjun{This portion is difficult to follow without having the figures alongside. Can we move the big study table closer and refer to it here?}\vidya{well, the big study table is for study 2 and does not show the sketches that we describe here. Think it would be worthwhile to include a sample set of sketches and refer to them below.}

\pheading{Cardinality \OnetoOne:}
From a total of 30 visualizations, there were 15 bar charts, 8 unit charts, 2 scatter plots, 2 pie charts, 1 line chart, and 1 box plot. 10 of the charts had a vertical orientation and 7 had a horizontal orientation. For the bar charts, seven were grouped bar charts and the rest were simple two-bar bar charts. We noticed a trend of both the values being in a different color. The simple bar charts were annotated with the value of the attribute on top of the bar. There were three adjacent labeled unit charts where the data values were mentioned in a bigger font below the circles, and there were four other unit charts. There was a trend of values included in the charts as legends or labels.
For \OnetoOne, we created (A) a simple bar chart, (B) an adjacent unit chart, (C) a horizontal unit chart, and (D) a grouped bar chart.

\pheading{Cardinality \OnetoN:}
From a total of 35 visualizations, there were 21 bar charts, 5 scatterplots, 4 dot plots, 2 unit charts, 2 line charts, and 1 pictograph. 12 of the charts had a horizontal orientation. For the bar charts, 2 were grouped bar charts and 2 stacked. Generally, there were multi-bar bar charts and two-bar bars charts where one bar depicts a value and the other depicts the average value of others. We also noticed a general trend of coloring the value mentioned in the query in a different color as compared to others. The simple bar charts were annotated with the value of the attribute on top of the bar. There was another trend of data values included in the charts as legends or labels. 
For \OnetoN, we created (E) a horizontal multi-bar chart, (F) a horizontal simple bar chart, (G) a dot plot, and (H) a scatterplot. 

\pheading{Cardinality \N:}
From a total of 31 visualizations, there were 17 bar charts, 5 scatterplots, 3 box plots, 1 line chart, 1 parallel coordinates chart, and 1 sentiment chart. 10 of the charts had a horizontal orientation and 9 had a horizontal orientation. For bar charts, there were four small multiples, five stacked, and two grouped. We also noticed a general trend of all the values being in the same color representing a single attribute in the case of bar charts. All the values were shown in different colors or depicted in a legend in the case of scatter plots. 
For \N, we created (I) a horizontal multi-bar chart, (J) a small multiple bar chart, (K) a scatterplot, and (L) a box-plot.

\pheading{Cardinality \NtoN:}
From a total of 41 visualizations, there were 21 bar charts, 5 scatterplots, 6 dot plots, 2 box plots, 2 unit charts, 4 line charts, and 1 pie chart. 15 of the charts had a vertical orientation and 10 had a horizontal orientation. For bar charts, six were grouped bar charts, five small multiples, and two stacked. We also noticed a general trend of all the values of n and all the values of m in a different color. All the values were in a different color in the case of scatterplots and grouped bar charts.
For \NtoN, we created (M) a grouped bar chart, (N) a small multiples bar chart, (O) a simple vertical bar chart, and (P) a scatterplot.

\section{Study 2: Identifying Visualization Preferences}
\label{sec:study2}
We conducted an online crowdsourced experiment to cross-validate user preferences for the 16 representative visualizations identified in Section~\ref{ExtractedVisualizations} for our design space. 
We focus on non-expert user preferences because the designs from visualization experts are often meant for communicating key patterns to naive viewers to help them make comparisons. 
The results of this study can inform the design choices for visualization recommendation systems focused on comparisons. 

\subsection{Participants}
Based on a pilot study with 10 participants using only \OnetoOne\ queries, we conducted a power analysis to determine the number of participants required for the experiment to find an overall difference in preference rankings. With a medium effect size of $\sim$0.49, our analysis suggests that a target sample of 76 would yield 95\% power to detect an overall difference between preference rankings for the four visualizations at an alpha level of 0.05. 
We recruited 79 participants via Prolific.com \cite{palan2018prolific} to complete an online survey through Qualtrics \cite{qualtrics2013qualtrics}. They were compensated at 10.55 USD per hour. In order to participate in our study, the workers had to be based in the United States and fluent in English. After excluding participants who failed attention checks (e.g., failing to select a specific answer in a multiple-choice question) or entered illegible/nonsensical response, we ended up with 77 participants, with 58 that identified as women ($M_{age}$ = 39.34, $SD_{age}$ = 16.18), 18 as men ($M_{age}$ = 40.94, $SD_{age}$ = 13.90), and one chose to not disclose.  

The participants completed a subjective graph literacy report~\cite{garcia2013communicating} and reported an average value of 3.87 out of 6 ($SD = 0.82$, 1 = low self-reported literacy, 6 = high self-reported literacy), suggesting that most participants were comfortable with visualizations but did not identify as visualization experts. Only 5 people reported that they create visualizations often for work or as a hobby, and 17 people reported that they rarely interact with visualizations in their day-to-day life. 

We asked our participants how much effort they put in completing our study; they were told that the answer to this question would not affect their compensation and were encouraged to answer honestly. 56 participants reported having put in a lot of effort into our study, carefully thinking through their answers before responding.~21 participants reported having put in some effort, having answered the questions without thinking too deeply about anything. 

\subsection{Stimulus and Design}
Similar to Study 1, we considered four cardinalities and four levels of concreteness, resulting in 16 comparison queries. We used the Amazon Books dataset \cite{saalu_2020} and generated queries similar to those from Study 1. The dataset had seven attributes including three categorical (e.g., Book Title, Author), three numeric (e.g., User rating, Reviews), and one temporal attribute (Year). We created visualizations following the representative designs extracted from Study 1 (Section~\ref{ExtractedVisualizations}). We varied the data attributes and values used to generate the visualization to match the content of the queries for each cardinality. 
For the 16 queries, we made 64 visualizations using Vega-Lite \cite{10.1109/TVCG.2016.2599030}; 16 visualizations for each cardinality that included four visualizations for each level of concreteness (query).
We briefly describe how the visualizations looked for each level of concreteness in Study 2. The full set of comparison queries and example visualizations can be found in Figure~\ref{Exp2OverallResults}.

\pheading{\EVEA:} For \OnetoOne\ (charts A, B, C), \OnetoN\ (E, F, G), \N\ (I, L), and \NtoN\ (M, N, O) the visualizations were straightforward with explicit data values compared to an explicit data attribute (\texttt{User rating}) in case of \OnetoOne\ and the data attribute (\texttt{Price}) in case of \OnetoN , \N , and \NtoN . 

\pheading{\EVIA:} For \OnetoOne\ (charts A, B, and C), \OnetoN\ (E, F, and G), \N\ (I, and L), and \NtoN\ (M, N, and O) the visualizations were straightforward with explicit data values compared to an implicit data attribute (``Popularity''), which is a vague modifier and is mapped to the attribute (\texttt{Reviews}). 

\pheading{\IVEA:} For \OnetoOne\ (charts A, B, and C), \OnetoN\ (E, F, and G), \N\ (I, and L), and \NtoN\ (M, N, and O) the visualizations were straightforward with implicit data values compared to an explicit data attribute (\texttt{User rating}) in case of \OnetoOne\ and the attribute (\texttt{Price}) in case of \OnetoN, \N, and \NtoN. The titles of implicit values such as ``a bestseller book in 2012'' in case of \OnetoOne , ``high rated books'' in case of \OnetoN\ and \N , and ``high rated fiction books'' in case of \NtoN\ are shown in the charts as labels or legends.

\pheading{\IVIA:} For \OnetoOne\ (charts A, B, and C), \OnetoN\ (E, F, and G), \N\ (I, and L), and \NtoN\ (M, N, and O) the visualizations show implicit data values compared to an implicit data attribute (``Popularity'') which is mapped to the attribute (\texttt{Reviews}). The titles of implicit values such as ``a cheap book'' in case of \OnetoOne\ and \OnetoN , ``expensive books'' in case of \N , and ``high rated non fiction books'' in case of \NtoN\ are shown in the charts as labels or legends.

%%%%%%%%%%%%%%%%%%%%%%
\subsection{Procedure}
Participants were given a link to our survey.
After consenting to participate, participants were given a metadata sheet describing the variables in the Amazon Books dataset. Participants had access to this metadata sheet throughout the experiment. 
After a brief introduction to the survey, participants viewed all sixteen comparison queries in random order on separate pages.
For each comparison query, participants were given the four corresponding visualizations based on the cardinality of the query and were instructed to rank them in terms of how well they enable a viewer to make the intended comparison.
Participants were told to assume that the viewer was not familiar with the dataset and there were no correct answers. 
At the end of the survey, participants reported demographic information, completed the self-report visual literacy test, and reported the amount of effort they put in completing the study.

%%%%%%%%%%%%%%%%%%%%%%%%%%%%%%
\subsection{Results for Each Comparison Utterance}
\begin{figure*}[t!]
\centering
 \includegraphics[width = \linewidth]{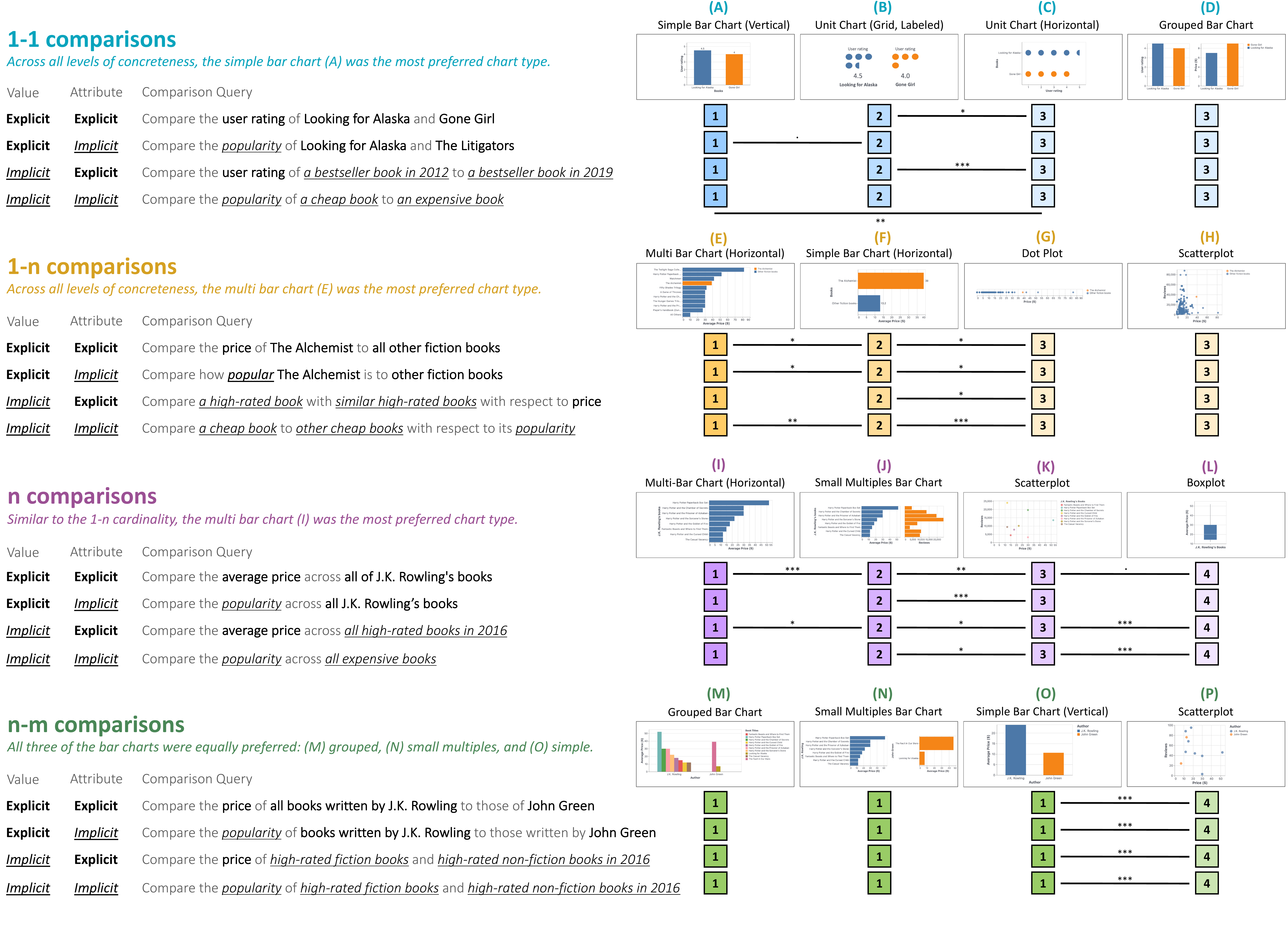}
 \vspace{-25px}
 \caption{Study 2 preference rankings for visualizations extracted based on Study 1 (Section~\ref{ExtractedVisualizations}). Each row represents a query with varying cardinalities and levels of concreteness (\textbf{explicit} or \emph{\underline{implicit}}). The boxed number represents the average preference ranks assigned to each visualization, based on a Friedman Rank Sum test. Connecting lines represent significant differences based on post-hoc comparisons. Three asterisks (***) indicate that the comparison p-value $<$ 0.001. Two asterisks (**) represent p $<$ 0.01. One asterisk (*) represents p $<$  0.05. One dot (.) represents p $<$  0.1.}
 \label{Exp2OverallResults}
\vspace{-1.25em}
\end{figure*}

We conducted a Friedman Rank Sum test for each comparison query using the PMCMRplus R package~\cite{pohlert2018package} to compare preference rankings for the four visualizations, with post-hoc pair-wise comparisons via Conover's test with Bonferroni's correction to determine the specific ranking differences (Table with complete analysis can be found in the \href{https://osf.io/zp6kh/?view_only=dfcf6012d3654fd99884ae8523e4f1aa}{supplementary material}). Figure~\ref{Exp2OverallResults} shows the summary results. 

%\noindent\cx{Need to replace ABCD with the appropriate imp/exp, see figure for details}...

% \input{tables/study2-results}

\pheading{Cardinality \OnetoOne:} Accounting for all combinations of concreteness, participants generally preferred (A)~\textcolor{oneToOneColor}{simple bar charts} or (B)~\textcolor{oneToOneColor}{adjacent unit charts} to help them make \OnetoOne\ comparisons. They preferred (C)~horizontal unit charts and (D)~grouped bar charts significantly less. 
There is a significant difference in user preferences for the \EVEA\ query (Friedman $\chi^2$ = 57.28, $p < 0.001$), the \EVIA\ query (Friedman $\chi^2$ = 21.85, $p < 0.001$), the \IVEA\ query (Friedman $\chi^2$ = 80.97, $p < 0.001$), as well as the \IVIA\ query  (Friedman $\chi^2$ = 18.29, $p < 0.001$).

%%%%%%%%%%%% 1-n
\pheading{Cardinality \OnetoN:}
Accounting for all combinations of implicit and explicit data variables and data attributes, participants generally preferred (E)~\textcolor{oneToNColor}{multi bar charts} to help them make \OnetoN\ comparisons. Participants preferred (G)~dot plots and (H)~scatterplots significantly less.
There is a significant difference in user preferences for the \EVEA\ query (Friedman $\chi^2$ = 71.42, $p < 0.001$), the \EVIA\ query (Friedman $\chi^2$ = 60.45, $p < 0.001$), the \IVEA\ query (Friedman $\chi^2$ = 91.51, $p < 0.001$), as well as the \IVIA\ query  (Friedman $\chi^2$ = 88.51, p $<$ 0.001).

%%%%%%%%%%%% n
\pheading{Cardinality \N:}
Accounting for all combinations of implicit and explicit data variables and data attributes, participants generally preferred (I)~\textcolor{nColor}{multi bar charts} to help them make \N\ comparisons, similar to the \OnetoN\ cardinality. Participants preferred (L)~box plots the least, with (J)~small multiples bar charts and (K)~scatterplots in the middle of the pack.
There is a significant difference in user preferences for the \EVEA\ query (Friedman $\chi^2$ = 123.34, $p < 0.001$), the \EVIA\ query (Friedman $\chi^2$ = 123.34, $p < 0.001$), as well as both the \IVEA\ and \IVIA\ queries (Friedman $\chi^2$ = 113.83, $p < 0.001$ for both queries).
%, as well as the \IVIA\ query (Friedman $\chi^2$ = 113.83 (same as previous, not a mistake), $p < 0.001$).

%%%%%%%%%%%% n-m
\pheading{Cardinality \NtoN:}
Accounting for all combinations of concreteness, participants equally preferred (M)~\textcolor{nToMColor}{grouped bar charts}, (N)~\textcolor{nToMColor}{small multiples bar charts}, and (O)~\textcolor{nToMColor}{simple bar charts} to help them make \NtoN\ comparisons. Participants generally preferred (P)~scatterplots the least. 
There is a significant difference in user preferences for the \EVEA\ query (Friedman $\chi^2$ = 58.24, $p < 0.001$), the \EVIA\ query (Friedman $\chi^2$ = 69.36, $p < 0.001$), the \IVEA\ query (Friedman $\chi^2$ = 76.78, $p < 0.001$), as well as the \IVIA\ query (Friedman $\chi^2$ = 92.38, $p < 0.001$).

\subsection{Summary Findings} 
Results from Study 2 provide insights into which visualization arrangement may be preferred for a given type of comparison utterance from our design space. Overall, we found that the preference ranking of charts within the four cardinalities (i.e., \OnetoOne, \OnetoN, \N, and \NtoN) was consistent, even taking into account the combinations of concreteness (Figure~\ref{Exp2OverallResults}). We use the notation [$W\#$] when referring to participants in these studies.
We now discuss specific preferences and participant feedback for the various cardinalities of comparison utterances.
\vspace{-1mm}

\pheading{\OnetoOne\ Comparisons:} Viewers preferred the (A)~\textcolor{oneToOneColor}{simple bar charts} without the need for unit charts or grouped bar charts: \emph{``The bar graphs are easier to compare than the bubbles.''}~[$W18$] and \emph{``The simple bar plot was easiest to understand and provided a perfect conveyance of the data.''}~[$W33$]. Other viewers had trouble interpreting unit charts; for example, \emph{``I have no idea what [the unit chart] is trying to say.''} [$W12$].

\pheading{\OnetoN\ Comparisons:} Viewers preferred (E)~\textcolor{oneToNColor}{multi-bar charts} sorted in descending order with the singleton value in the comparison utterance highlighted for easier relative judgment tasks. As a second option, people indicated a preference for (F)~\textcolor{oneToNColor}{a simple horizontal bar chart} comparing the singleton value to an aggregated bar of other entities: \emph{``The first one [multi-bar chart] is the best visual as it gives the most accurate ranking system, while the other one [simple horizontal bar chart] only shows averages, which could cause misrepresentation as it lumps all the books.''} [$W68$]. (G)~Dot plots and (H)~scatterplots were both equally ranked lower in the rankings. $W65$ summarized that \emph{``The bar chart is easy to compare due to The Alchemist having a brighter color that stands out on the chart. The scatterplot is somewhat confusing due to the cluster of circles bunched together.''}

\pheading{\N\ Comparisons:} The top two user preferences were (I)~\textcolor{nColor}{a multi-bar chart} and (J)~\textcolor{nColor}{small multiples of simple bar charts} showing the entities sorted by a common property such as the \texttt{price}. Participants ranked the (K)~scatterplot lower than the bar charts with the (L)~box plot ranked last. $W5$ commented that \emph{``The first two graphs [bar graphs] are just as easy to understand and the third graph [scatterplot] is a little less easy to understand and the last one [box plot] is objectively pretty terrible at illustrating the popularity across all of JK Rowling's books.''}

\pheading{\NtoN\ Comparisons:} There was a tie in preferences among the three types of bar charts: (M)~\textcolor{nToMColor}{grouped}, (N)~\textcolor{nToMColor}{small multiples}, and (O)~\textcolor{nToMColor}{a simple bar chart}. (P)~Scatterplots were consistently ranked last. Viewers liked to see the breakdown of the entities being compared rather than viewing aggregated data: \emph{``The first two choices breakdown the price by book, while the last two choices don't have that breakdown.''} [$W2$]. Several participants expressed difficulty interpreting the scatterplot for a comparison task; for example, $W63$ noted that
\emph{``Scatterplots are difficult to read. The first option is cleanest and easiest to read to compare the cost of books by the author.''} and $W25$ explained in detail that \emph{``My first choice showed at a quick glance the exact comparison between data sets - and number two was relatively easy as well as three. But number four, again... those dots!''}

\section{Discussion}

To summarize, the findings from both studies are consistent with that from prior work, where bar charts are generally conducive to visual comparisons~\cite{xiong2021visual}. However, we found interesting insights regarding the type of bar charts that were preferred. Feedback from participants showed that bars were easily comparable when visually aligned and spatially proximate. In terms of language, we observed a trend towards requiring interactivity and annotations where implicit values/attributes were involved. Insightful trends were found regarding  the general interpretation of vague modifiers and underspecified implicit values by the participants. Observations from these studies provide four key design implications for recommendation tools and NLIs to support comparisons during visual analysis.

\subsection{Design Implications}

\pheading{Basic charts are reasonable responses for a variety of comparisons}. The preference ranking consistently showed that bar charts are preferred for their simplicity in representing comparisons. Our findings suggest that for \OnetoOne, vertical bar charts and unit charts were preferred. For \OnetoN, preference was towards horizontal bar charts. Horizontal and small multiple bar charts were popular for \N; whereas for \NtoN, grouped, small multiple or simple bar charts were preferred. Many tools are already capable of creating basic charts such as sorted horizontal and vertical bar charts with support for highlighting a subset of bars for easier comparison. By extending the language parser in these tools to recognize comparison expressions, they can support a repertoire of comparison types without the need to generate bespoke chart types.

\pheading{Include necessary information useful for the comparison task.} Viewers ranked charts that contained only information relevant to the comparison higher than those that had extra or unnecessary information, an observation synergistic with previous NLI research for visual analysis tasks~\cite{hearst2019toward,chatbotsetlur:2022}. 
When generating visualization responses, tools should ensure that they contain only information that is indicated in the comparison utterances. Participants explained their rationale for ranking charts with superfluous information lower than others: 
\emph{``The bottom chart [small multiples bar chart] contains extraneous information that is not needed for the comparison sentence (price).''}~[$W6$]. Similarly, for comparisons involving specific values, viewers preferred charts showing unaggregated data: \emph{``Preferred graph gives comparative information. All of the others compare one book with an aggregate.''}~[$W31$].

\pheading{Indicate how implicit entities are interpreted}. Emulating previous NL systems~\cite{hearst2019toward,sentifiers,eviza}, we map vague modifiers to more concrete representations. Viewers appreciate the importance of exposing the provenance of how implicit entities such as ``cheap'' and ``best-selling'' are mapped to specific data attributes in the visualization. $W72$ remarked that \emph{``The first option [small multiples bar chart] allowed readers to gauge both price and review at the same time, while the other graphs did not show them at all.~`Expensive' books are subjective''}. In general, for vague modifiers, participants usually picked all the attributes from the dataset that somewhat matched the implicit value or attribute and mapped all of them to it (e.g., `performance' as a combination of box-office revenue and IMDB ratings). Underspecified values were often interpreted as being above or below average.

\pheading{Support user interaction and text with the visualizations}. We found that users preferred interactivity (e.g., filter controls, changing the attribute, or modifying the data values), particularly for comparison utterances concerning implicit concepts. In addition, users preferred the inclusion of text describing how these implicit concepts were interpreted in the visualization responses. As shown in Table 1, \IVEA~had the highest percentage of visualizations with interactive features, closely followed by \IVIA. \IVEA~also had the highest percentage of visualizations with annotations, closely followed by \EVIA~and \IVIA. In terms of cardinality, \N~had the highest percentage of visualizations with interactive features, followed by \OnetoN. \N~also had the highest number of visualizations with annotations, followed by \NtoN.

\subsection{Limitations and Future Directions}
Our work specifically explores the interplay between language and visual representations for comparisons. However, we discuss some limitations of our work and identify promising future directions in this important area of research.

\pheading{Explore additional design considerations for comparisons.} The visualizations used in Study 2 varied across several design dimensions (e.g., arrangement, encodings, chart type, annotation), unlike prior studies~\cite{xiong2021visual, xiong2021grouping}), where one design element (such as visual arrangement) was isolated and tested comprehensively. 
We considered the current approach because past work has shed light on how vast and complex the experimental space can be. 
To avoid a combinatorial explosion of experimental conditions, we hence adopted a design-and-then-validate approach to cover a wide range of visualization designs for multiple types of comparison utterances.
The approach enabled us to produce immediately actionable guidelines for visualization recommender and NL systems. 
Future work should examine how interactivity, such as brushing, linking, dynamic filtering and scaling, and annotation techniques might facilitate different types of comparisons.
 
%As indicated from our findings, bar charts are a promising direction to facilitate comparisons, and future work can systematically test individual design decisions for bar charts, such as manipulating color, grouping, and layout choices to better support comparisons.

% Further, there are several design dimensions that we identified in Study 1 but did not investigate deeper in Study 2. Specifically, we did not test for the utility of interactivity and annotation in the visualizations, which by themselves would be separate research problems to pursue~\cite{hoffswell2020languages,stokesgive}. 

% \vidya{This para below does not belong here as it is mostly describing the set up and design of the study.}\\
% \textcolor{magenta}{Due to the remote nature of our study, we had to have a time limit of 1 hour for each participant to make visualizations for 4 utterances. However, participants were only expected to draw rough sketches of the visualizations and explain to us verbally in case they wanted to add extra details as we also realise that our participants were making the vis using a mouse. Furthermore, we also gave the participants some time before starting the study to explore the tool and get accustomed to it as well as with its shape assets.}
 
%VS: Updated.
%TODO: potential future research direction with regards to the ethics of recommendation systems.  
\pheading{Explore the role of ethics for interpreting comparison utterances.} With the impreciseness of language, we acknowledge that people may interpret implicit attributes and vague concepts in unique, subjective ways. There is a need to further explore the ethical issues surrounding how tools implementing comparison utterances can be transparent about their underlying assumptions and decision-making processes. Future work should specifically evaluate the ethical considerations for NLIs and recommendation systems~\cite{correll:2019,Zeng2022AnEF} for responsibly supporting analytical inquiry for comparisons.

\pheading{Leverage additional metrics beyond user preferences.}
%VS: Updated.
%\hl{TODO: discuss how other metrics in addition to user preference might help us design better NLI systems}
The visualization responses for comparison utterances were informed by a series of interviews that were subsequently validated by a user preference study. Future research should evaluate the effectiveness of these preferred visualizations by examining the speed and accuracy with which users can make the intended comparisons, or test the types of comparisons they afford via qualitative responses. 
Other research directions should explore how NLIs can take into account \emph{implicit} feedback by users through refinement and repair of system choices for visual comparisons as well as query reformulations~\cite{setlur:iui}. 
In addition, telemetry data collected from production systems and deployed visual analysis tools can be used as a training set for machine learning models to learn and improve user expectations over time. 
Other metrics involve \emph{explicit} feedback through surveys and a like/dislike functionality built into NLIs could provide additional insight into the utility of the visualization response in supporting a comparison intent.

\pheading{Extend comparisons to support exploratory data analysis.}
%VS: Updated.
%\hl{TODO: study doesn’t test exploratory analysis scenario, add to future directions to test this and also add to future directions}
Our paper focuses on a question-answering form of interaction for supporting comparison intents. However, future research should explore how users express comparisons to support exploratory data analysis, focusing specifically on language pragmatics for follow-up inquiry during an analytical conversation. Such insights can also be useful to help inform the design of recommender tools for suggesting comparison utterances and views as part of a larger analytical workflow.
 
\section{Conclusion}

Visual comparisons are an important form of analytical workflows as people reason about and make sense of data. With the growing popularity of NLIs and recommendation systems, interpreting and visualizing the semantic nuances of comparison utterances can be challenging. In this paper, we explore a preliminary design space of NL comparison utterances covering four cardinalities of comparisons and addressing varying degrees of concreteness in these comparison intents. Through a series of interviews with 16 visualization novices and experts, we create a potential mapping between comparison utterances in our design space and different visualization techniques. By empirically validating this mapping in a preference ranking study with 77 participants, we provide design implications for interpreting the comparison language for a range of scenarios. We hope that our findings are a step towards addressing the complex interplay between language and visualization.

% Reviewer 2 saying 'why didn't you test these design techniques systematically? why not conduct a separate study to test vertical/horizontal, different color choices, different arrangement and grouping etc."
% -> so we should probably motivate Exp 1 by saying "it's impossible to systematically test this via a series of controlled experiment. The number of condition would explode, and we would make very incremental steps. So we conducted the workshop with experts/non-experts, interviewed them and extracted key design features to systematically test in Exp 2.
% The authors have done a lot of incremental controlled experiments to tackle comparisons in visualization, such as comparing the effectiveness of spatial group to color grouping \cite{xiong2021perceptual}, or mapping the best visual arrangement to a given comparisons in bar charts \cite{xiong2021visual}. While controlled, incremental explorations in the design space is necessary, it's important to balance the realism with controlled to ensure external validity and to identify actionable guidelines that can be implemented and tested in real-world scenarios \cite{wall2022vishiker}

\acknowledgments{
The authors wish to thank Shubham Mishra and Nisarga Patil from UMass Amherst for helping with the coding process, the study participants for their feedback, and the reviewers for their thoughtful comments and suggestions.
% Adobe Research and Tableau Research for funding parts of this research project. 
}
\bibliographystyle{abbrv-doi}

\bibliography{references}
\end{document}